\documentclass{aa}
\usepackage{natbib}
\usepackage{graphicx,hyperref}
\usepackage{txfonts}
\usepackage[dvips]{color}
\newcommand{\aflux}{A$_\mathrm{flux}$}
\newcommand{\hi}{H\,{\sc i}}

\newcommand{\km}{km\,s$^{-1}$}

\newcommand{\prim}{$^{\prime}$}
\newcommand{\prin}{$^{\prime\prime}$}
\newcommand{\msolar}{M$_{\odot}$}
\newcommand{\lsolar}{L$_{\odot}$}

\newcommand{\degree}{$^{\circ}$}

\definecolor{gold}{rgb}{0.85,.66,0}
\begin{document}

   \title{\textcolor{black}{A} $\sim$\,12 kpc   \textcolor{black}{\hi\ extension} and other  H{\sc i} asymmetries  in the isolated  galaxy CIG\,340 (IC 2487)}

   \subtitle{}

   \author{T. C. Scott
           \inst{1,2} \fnmsep\thanks{ \email{tom@iaa.es}},
           C. Sengupta
          \inst{3},
         L. Verdes Montenegro
           \inst{1},
           A. Bosma
          \inst{4},
          E. Athanassoula
          \inst{4}, 
           \textcolor{black}{ J. Sulentic}
           \inst{1},        
           D. Espada
           \inst{5,6,7},
           M. S. Yun
           \inst{8},         
           \and
           \textcolor{black}{M. Argudo--Fern\'andez}
          \inst{1}
          }

\institute{Instituto de Astrof\'{i}sica de Andaluc\'{i}a (IAA/CSIC), Apdo. 3004, 18080 Granada, Spain \\
Centre for Astrophysics Research, University of Hertfordshire, College Lane, Hatfield, AL10 9AB, UK \\
Korea Astronomy and Space Science Institute, 776, Daedeokdae--ro, Yuseong--gu, Daejeon, 305-348, Republic of Korea\\
\textcolor{black}{Aix Marseille Universit\'e CNRS, LAM (Laboratoire d'Astrophysique 
de Marseille) UMR 7326, 13388, Marseille, France}\\
Joint ALMA Observatory (ALMA/ESO), Alonso de Córdova 3107, Vitacura, 763-0355, Santiago, Chile\\
National Astronomical Observatory of Japan (NAOJ), 2-21-1 Osawa, Mitaka, 181-8588, Tokyo, Japan\\
The Graduate University for Advanced Studies (SOKENDAI), 2-21-1 Osawa, Mitaka, 181-8588, Tokyo, Japan\\
Department of Astronomy, University of Massachusetts, 710 North Pleasant Street, Amherst, MA 01003, USA\\
}

   \date{Received ; accepted }

 
  \abstract
   { \textcolor{black}{\hi\ kinematic asymmetries are common in late--type galaxies irrespective of environment, although the amplitudes are strikingly low in isolated galaxies. As part of our studies of the \hi\ morphology and kinematics in isolated late--type galaxies we have chosen several very isolated galaxies from the AMIGA sample for \hi\  mapping. We present here the results of \hi\ mapping  of CIG\,340 (IC 2487) which was selected because \textcolor{black}{ its integrated \hi\  spectrum has a very symmetric } profile (\aflux\ = 1.03 $\pm$ 0.02).} }
 { \textcolor{black}{  \textcolor{black}{Optical images of the galaxy hinted at a  warped disk in contrast to the symmetric integrated \hi\ spectrum} profile.  Our aim \textcolor{black}{is}  to determine the extent to which the optical asymmetry \textcolor{black}{is} reflected in the resolved \hi\ morphology and kinematics.   }}
{\textcolor{black}{Resolved   21-cm \hi\ line mapping  \textcolor{black}{has been} carried out using the Giant Metrewave Radio Telescope (GMRT). The \hi\  morphology and kinematics from this mapping together with   other multi-wavelength data \textcolor{black}{have been} used to study the relationship between the \hi\ and stellar components of CIG\,340.  }}
{ \textcolor{black}{ GMRT observations reveal significant \hi\ morphological asymmetries in CIG\,340 despite it's \textcolor{black}{overall}  symmetric optical form and \textcolor{black}{highly} symmetric \hi\ spectrum. The most notable \hi\ features are: 1)  a  warp in the \hi\ disk (with an optical counterpart),  2) the \hi\ north/south flux ratio = 1.32 is much larger than  \textcolor{black}{expected from the integrated \hi\ spectrum} profile  and 3) a $\sim$ 45\prin\  (12 kpc) \textcolor{black}{\hi\ extension}, containing $\sim$ 6\% of the detected  \hi\ mass on the northern side of the disk.   }}
{  \textcolor{black}{ \textcolor{black}{Overall we conclude that in isolated galaxies a highly symmetric \hi\ spectrum can mask significant \hi\ morphological asymmetries which can be   revealed by \hi\ interferometric mapping.} The northern \textcolor{black}{\hi\ extension} appears to be the result of a recent perturbation (10$^8$ yr), possibly by a satellite which is now disrupted or projected within the disk.  But, we cannot rule out that the \textcolor{black}{\hi\ extension} and the other observed asymmetries  are the result of a} long lived dark matter halo asymmetry. This study provides  an important step in our ongoing program to determine the predominant source of \hi\ asymmetries in isolated galaxies.  For CIG\,340 the isolation from major companions, symmetric \hi\ spectrum, optical morphology and interaction timescales have allowed us to narrow the possible causes the \hi\ asymmetries and identify tests to further constrain the source of the asymmetries.}


\titlerunning{CIG\,340 \hi\ asymmetries}
\authorrunning{Scott T.C.  et al.}

\maketitle
%


\section{Introduction}
\label{intro}
\textcolor{black}{A  galaxy's evolution and its properties at \textit{z} = 0  are the result of
 both  internal processes and its environment.} \textcolor{black}{The AMIGA (Analysis
of the Interstellar Medium of Isolated GAlaxies (\url{http://amiga.iaa.es}) project's principal aim is  to provide \textcolor{black}{quantitative} benchmarks, from its   strictly defined sample of isolated galaxies \citep[the AMIGA sample,][]{vm05}, with  which to assess the impact of dense environments on galaxy properties. \textcolor{black}{The AMIGA project has clearly established that variables expected to be enhanced by interactions are lower in isolated galaxies than in any other sample}  \citep[Verley et al. 2007a,b;][]{lisen07,leon2008,sabater2008,durbala08,lisenfeld2011,fernadez12,fernadez13}. \nocite{verley07a} \nocite{verley07b}}
Galaxies in the AMIGA sample have remained free of  major tidal interactions  for \textcolor{black}{at least }the last $\sim$3 Gyr \citep{vm05} \textcolor{black}{and most spirals in the  sample likely host pseudo--bulges rather than classical bulges \citep{durbala08}}. 

\textcolor{black}{A  key finding  \textcolor{black}{from our studies}  of isolated galaxies was that they have smaller \hi\ spectral asymmetries than galaxies in denser environments \citep{espada11}. Espada et al. studied the \hi\ single dish spectral profiles of a sample of 166 AMIGA  galaxies using an \hi\ asymmetry parameter \aflux, defined as the ratio of the \hi\ flux between the receding and approaching sides of the spectrum. They \textcolor{black}{find that} the distribution of this parameter is well described by a half Gaussian distribution, with only 2\% of the sample having an asymmetry parameter in excess of 3 $\sigma$ (\aflux\ = 1.39), i.e., a flux excess $>$ 39\% in one half of the spectrum. In contrast, field galaxy samples deviate from a Gaussian distribution and have a higher fraction (10-20\%) of asymmetric galaxies \citep{espada11}.}

It is well known that the environment affects 
galaxy evolution: tidal interactions can perturb both the stellar
and gas disks, ram pressure is able to perturb \hi\ disks \textcolor{black}{
\citep{voll08} \textcolor{black}{and} major mergers can destroy the structure of both the gas and stellar
disks} \citep{struck99}.  The presence of asymmetric \hi\ profiles in
isolated galaxies and the Gaussian distribution of their
\aflux\ parameters \citep{espada11} implies that  
process(es) other than interactions \textcolor{black}{with major companions} are operating to  maintain  long lived or frequent \textcolor{black}{shorter lived} perturbations in
isolated late--type galaxies.

\textcolor{black}{Several secular perturbation processes  have been proposed as the cause of \textcolor{black}{the observed} \hi\ and stellar asymmetries in isolated galaxies}:  accretion of cold gas from the surrounding environment, intermittent accretion of satellite galaxies and internal bar, disk or retarded SF driven perturbations \citep{bergvall95,bournaud02,bournaud05,sancisi08}. To date   very few  detailed observational  studies have been carried out   to determine the causes of such asymmetries in isolated galaxies, \textcolor{black}{although the \cite{pisano02} \hi\ study found $\sim$ 25\% of their isolated galaxy sample had gas rich minor companions.}

We are therefore carrying out a programme of resolved \hi\ studies for a sample of AMIGA galaxies to determine the \textcolor{black}{dominant  mechanism(s) that gives rise to the observed \hi\ spectral} asymmetries in isolated galaxies.   The \aflux\ spectral asymmetry parameters of the AMIGA galaxies for which we have carried out resolved \hi\ studies \textcolor{black}{so far, i.e., CIG\,96 \citep{espada05,espada11b}, CIG\,292 \citep{portas} and CIG\,85 \citep{sengupta12},  are  1.16, 1.23 and 1.27} respectively.   As part of this study  we present  \textcolor{black}{Giant Metrewave Radio Telescope  (GMRT) interferometric} \hi\ observations of the AMIGA late--type galaxy CIG\,340, which has one of the most symmetric single dish \hi\ \textcolor{black}{spectra}  in the \cite{espada11} AMIGA sample. Its \aflux\ = 1.03 $\pm$ 0.02  \textcolor{black}{from} the single dish spectrum  raised the expectation  \textcolor{black}{of an \hi\  disk with \textcolor{black}{a highly symmetric morphology at high angular
            resolution
}.} 

\begin{figure*}
\centering
\includegraphics[scale=0.62]{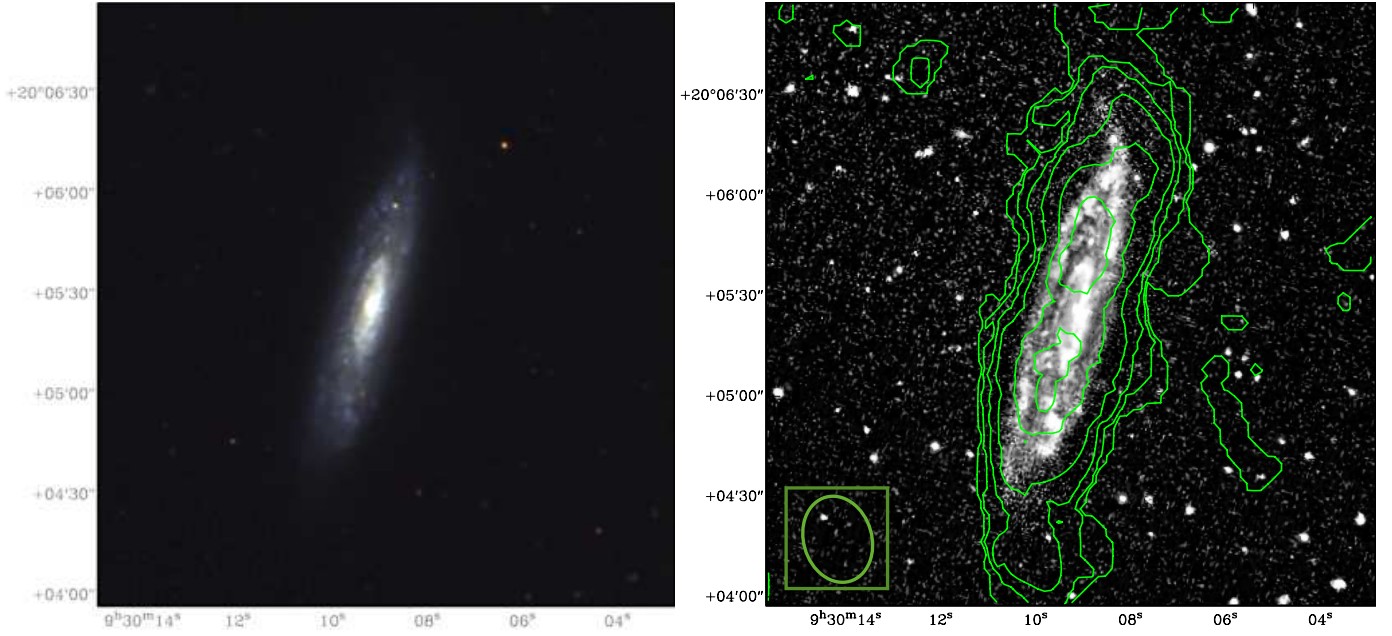}  
\caption{\textbf{CIG\,340 \textit{Left}}: Composite \textcolor{black}{SDSS \textit{ g, r, i} -- band false colour  image. \textbf{\textit{Right:}} \textcolor{black}{ A greyscale unsharp masked version of the image on the left, highlights the  faint disk edge details and is over plotted with  contours from the high resolution integrated \hi\ map. Details of the \hi\ contour levels are given in the caption for  the high resolution map in Figure \ref{mom0} .  }}}.
\label{sdss}
\end{figure*}

CIG\,340 (Figure \ref{sdss}) is a mid--sized spiral with  an AMIGA \textcolor{black}{RC3\footnote{Third Reference Catalog of Bright Galaxies.} morphology classification  of}  \textcolor{black}{ 4 (Sbc), and both its L$_{FIR}$ \citep{lisen07} and SDSS \textit{g -- r} colour  \citep{fernadez12} are consistent with a Sbc spiral but it is over--luminous in L$_B$ for an Sbc galaxy.  \cite{hernadez08}   classify CIG\,340 as an Sc spiral.  } Table \ref{table_a} gives some  further properties of the galaxy.

The  highly symmetric  \hi\ spectrum  and other properties  typical of AMIGA galaxies makes this well isolated  galaxy a particularly good candidate for  an \hi\ morphology study.  This  paper presents the results of our  GMRT  CIG\,340 observations as well as  utilising SDSS\footnote{Sloan Digital Sky Survey.}  \textcolor{black}{and \textit{GALEX}\footnote{The Galaxy Evolution Explorer.} } public archive images. \textcolor{black}{Section \ref{obs} sets out details of our observations, and the results are given in section \ref{results}. We briefly discuss} the \textcolor{black}{isolation of CIG\,340 and the} mechanisms which may be responsible for the observed \hi\  asymmetries in section \ref{discuss}. Concluding remarks are set out in section \ref{conclusions}.  J2000 coordinates are used throughout the paper, \textcolor{black}{ including the figure coordinates}.

\begin{table}
\centering
\begin{minipage}{190mm}
\caption{CIG\,340 parameters}
\label{table_a}
\begin{tabular}{l@{}l@{}l}
\hline
\textbf{\textcolor{black}{P}roperty}&\textbf{\textcolor{black}{V}alue} &\textbf{ \textcolor{black}{R}eference } \\ 
\hline
V$_{optical}$&\textcolor{black}{4339 $\pm$ 6 \,\km} &\textcolor{black}{\cite{fernadez12} }\\
RA&09h30m09.09s&Leon \& Verdes-\\
  & & Montenegro (2003) \nocite{leon03}\\
DEC&+20d05m24.3s& \hspace{.1cm}" \hspace{1.5cm}" \\
Distance&54.8 Mpc& \cite{fernadez12} \\
Spatial scale&$\sim$15.9 kpc/arcmin& \hspace{1cm}" \hspace{1.5cm}" \\
\textcolor{black}{D$_{25}$ major /minor}&  1.84 x 0.46 arcmin& \hspace{1cm}" \hspace{1.5cm}" \\
\textcolor{black}{D$_{25}$ major /minor}& \textcolor{black}{ 29.3 x 7.4 kpc}& \hspace{1cm}" \hspace{1.5cm}" \\
\textcolor{black}{Inclination}& \textcolor{black}{ 83.7\degree }& \hspace{1cm}" \hspace{1.5cm}" \\
Morphology&\textcolor{black}{4 (Sbc)}& \hspace{1cm}" \hspace{1.5cm}"  \\
\textcolor{black}{\hi\ \aflux}& \textcolor{black}{ 1.033 \textcolor{black}{$\pm$0.02}}& \textcolor{black}{\cite{espada11} }\\
log($L_B$)& 10.23  \lsolar &\textcolor{black}{\cite{fernadez12} }  \\
\textcolor{black}{B$_T^c$}& \textcolor{black}{12.989 }   & \hspace{1cm}" \hspace{1.5cm}"   \\
\textcolor{black}{log($M_*$) }&10.36 \msolar & \cite{fernadez13}  \\
log(L$_{FIR}$) & \textcolor{black}{9.53}\footnote{\textcolor{black}{From \cite{lisen07} adjusted for currently estimated distance.}} \lsolar& \textcolor{black}{\cite{lisen07} }\\
\hline
\end{tabular}
\end{minipage}
\end{table}

\section{Observations}
\label{obs}
 21--cm \hi\ line  emission from CIG\,340  \textcolor{black}{has been} observed for \textcolor{black}{10} hours with the GMRT in November, 2009. 
The full width at half maximum (FWHM) of the GMRT primary beam at  \textcolor{black}{1.4} GHz is $\sim$24\arcmin. 
The baseband bandwidth used  for  the observations was 8 MHz giving a
velocity resolution of $\sim$13.7 \km\ \textcolor{black}{within} the velocity range \textcolor{black}{3650 \km\ to 5150 \km.} Observational parameters, including the rms noise and beam sizes used to produce the integrated \hi\ maps are  presented in Table \ref{table2}.

Our  GMRT data  was reduced using the Astronomical Image Processing System (\textsc{AIPS}) software package. 
Bad data due to malfunctioning antennas, antennas with abnormally low gain and/or  radio frequency interference (RFI)
were flagged.  The primary flux density calibrator used in the observations was \textcolor{black}{3C147, and  the phase calibrators were 0842+185 and 1111+199}.  The flux densities are on the scale
from \cite{baars77}, with flux density uncertainties of $\sim$5 per cent. \textcolor{black}{Continuum subtraction in the uv domain was carried out with the \textsc{AIPS} tasks \textsc{uvsub} and \textsc{uvlin}. The continuum subtracted uv  data was transformed into the image plane using the  \textsc{AIPS}  task  \textsc{imagr}  and integrated \hi\ and  \hi\  velocity field maps were then extracted using the task \textsc{momnt}. }To analyse the \hi\  structures in CIG\,340 we produced image cubes and maps at different resolutions by tapering the
data with different uv limits. For this paper we  present maps  from \textcolor{black}{ cubes} with 
beam sizes of \textcolor{black}{ 26.08$^{\prime\prime}$  x 19.97$^{\prime\prime}$ (high resolution)  and 45.57$^{\prime\prime}$  x 41.35$^{\prime\prime}$ (low resolution). It should} be noted that the shortest
spacing between GMRT antennas is $\sim$ 60 m which in L band
\textcolor{black}{ results in  the observations being insensitive to smooth structures with angular sizes $>$ 6$^{\prime}$  to 7$^{\prime}$ ($\sim$ 100 kpc), i.e. is of little concern for  the observed source.}

\begin{table*}
\centering
\begin{minipage}{130mm}
\caption{GMRT \hi\ observation details}
\label{table2}
\begin{tabular}{lrrr}
\hline
&&\textcolor{black}{H}igh&\textcolor{black}{L}ow\\
&&resolution& resolution\\
&&cube& cube\\
\hline

Observation Date &13 November 2009& \\
Primary Calibrator & 3C147 & \\
Phase Calibrators & \textcolor{black}{0842+185, 1111+199}& \\
Phase Calibrator -- flux density [Jy]&\textcolor{black}{1.4, 1.3 } \\
Integration time [hr]& 10 &  \\
\textcolor{black}{Channel width [\km] }&&\textcolor{black}{27}&\textcolor{black}{13}\\
rms (per channel) [ mJy beam$^{-1}$]&& \textcolor{black}{1.0} &\textcolor{black}{1.4}\\
Beam (major axis)[$^{\prime\prime}$] &&26.08& 45.57 \\
Beam (minor axis)[$^{\prime\prime}$] &&19.97 &41.35 \\
\textcolor{black}{Beam PA} [\degree] &&15.5& -59.73 \\

\hline
\end{tabular}
\end{minipage}
\end{table*}

\section{Observational Results}
\label{results}

\begin{figure*}
\centering
\includegraphics[scale=0.65]{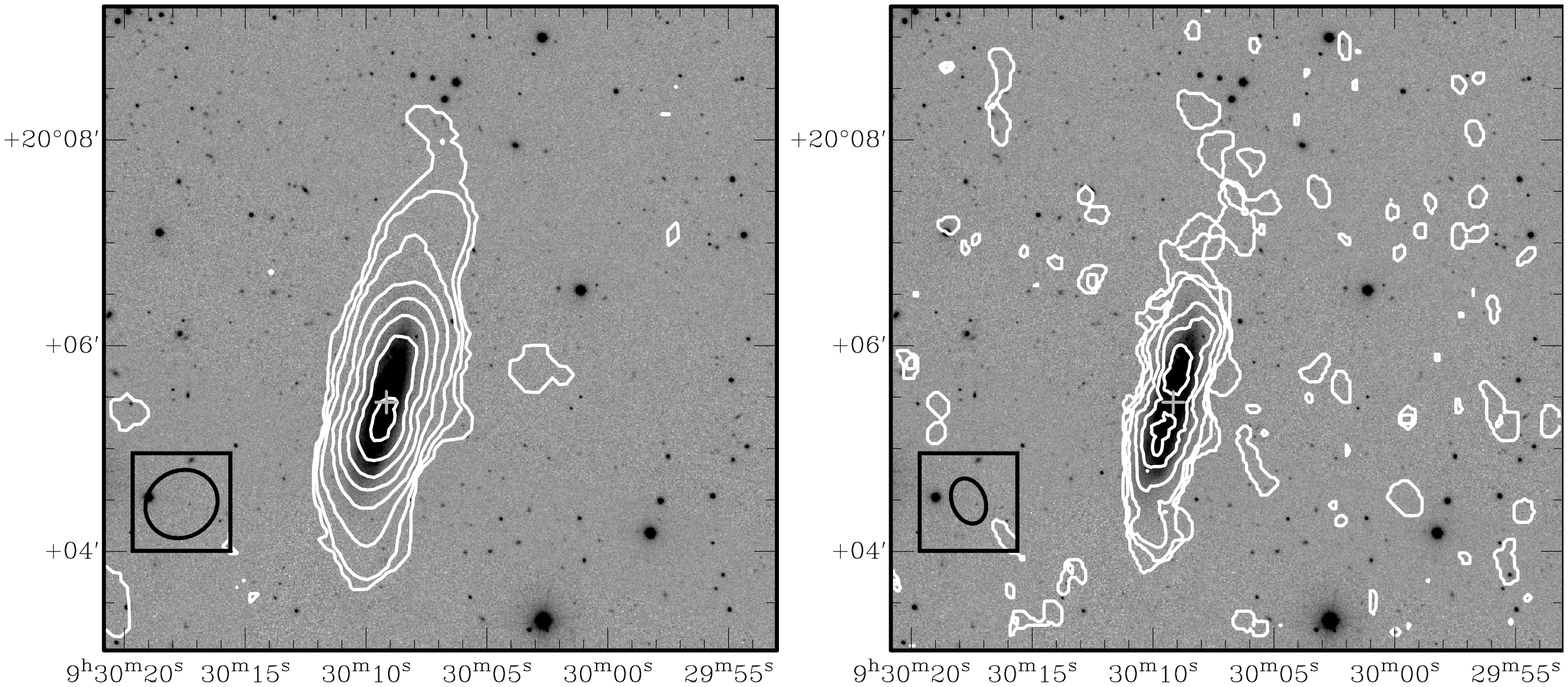}  
\caption{ Integrated  \hi\ \textcolor{black}{emission contours (white)} \textcolor{black}{overplotted} on a SDSS \textit{r}--band images. \textit{(\textbf{Left:}) }\textcolor{black}{ Integrated emission contours from the low resolution  \hi\ map,} (beam size = \textcolor{black}{45.57$^{\prime\prime}$ ′ x 41.35$^{\prime\prime}$)} where the \hi\ column
density levels are \textcolor{black}{10$^{20}$ atoms cm$^{−2}$ x (0.5, 1.2, 2.3, 4.7, 7.0, 9.3, 11.6, 14.0)}.  \textit{(\textbf{Right:})} \textcolor{black}{ Integrated emission contours from the high resolution  \hi\ map, }(beam size = \textcolor{black}{26.08$^{\prime\prime}$  x 19.97$^{\prime\prime}$}) where the
the \hi\ column
density levels are 10$^{20}$ atoms cm$^{−2}$  x \textcolor{black}{ (1.6, 3.2, 6.3, 10.6, 19.0, 25.3).}
\textcolor{black}{The first contours are at the 3 $\sigma$ level.  \textcolor{black}{At the bottom left of each panel a black ellipse shows the beam size.} The optical centre of the galaxy is marked with a  grey  cross.} }
\label{mom0}
\end{figure*}

\subsection {HI morphology and content }
\label{hi_content}

Figure \ref{mom0}  \textcolor{black}{ shows the contours from  the  GMRT low and high resolution   integrated  \hi\ maps overlaid on an SDSS \textit{r}--band image. The low resolution ($\sim$ 45$^{\prime\prime}$) \hi\ map contours (left panel) show the \hi\ extending  170\prin\ (45 kpc) north of the optical centre but only  125\prin\ (33 kpc) to the south, as measured at the 3 $\sigma$ noise level (a column density of 5 x 10$^{19}$  atoms cm$^{-2}$). The high resolution map -- right panel ($\sim$ 25$^{\prime\prime}$)  also   shows \textcolor{black}{an}  \textcolor{black}{ an ``S" shape extension continuing}  $\sim$ 44$^{\prime\prime}$ (12 kpc)  from northern edge of the optical disk, although with a clumpier morphology than at low resolution.  In both  the high and low resolution maps this  \textcolor{black}{\hi\ extension is projected } $\sim$ 45 $^{\prime\prime}$ (12 kpc) further north of the optical centre than the \hi\ extends in the south.  We use this \textcolor{black}{north/south} extent difference as an estimate of the \textcolor{black}{ \hi\ extension's} length and \textcolor{black}{its \hi\ mass} is  $\sim$ 4 x 10$^8$ \msolar, i.e., $\sim$ 6\% of the total \hi\ mass. This mass is about twice the  \hi\ mass asymmetry implied from the \aflux\ = 1.03 $\pm$ 0.02,  \textcolor{black}{although within the order of the  \hi\ mass inferred from the  \aflux\ uncertainty}.   The high resolution  integrated \hi\ map contours, Figure \ref{mom0} -- right panel,  shows \textcolor{black}{that} the  highest  \hi\  column densities \textcolor{black}{($N_{HI}$), $\geq$ 2.5 x 10$^{21}$ atoms cm$^{-2}$,} are located   in two maxima  $\sim$ 12$^{\prime\prime}$  (3.1 kpc) south and $\sim$ 17$^{\prime\prime}$  (4.5 kpc) north  of the optical centre. \textcolor{black}{$N_{HI}$ [atoms cm$^{-2}$]  is calculated from the equation:
\begin{equation}
    N_{HI} = 3.1 \times 10^{17} SdV /\theta^2
\end{equation}
Where,  $S dV$ = flux density [mJy beam$^{-1}$ \km] and $\theta$ = beam size [arcmin].}
 \textcolor{black}{The  right panel of Figure \ref{mom0}  shows the \hi\ column density at the optical centre is $\sim$ 1.9 x 10$^{21}$ atoms cm$^{−2}$, slightly less than at the maxima, but there is no clear} indication of an \hi\ hole at the galaxy center as is often the case for spirals with a strong classical bulge. }

The integrated \hi\ flux density obtained  by \cite{lewis85}  from single dish observations was 9.61 Jy \km\, which compares to 7.86  Jy \km\ from our the GMRT observations, i.e., $\sim$ 82\% of single dish  \hi\ flux density.  Conversion of the single dish  flux density to an \hi\ mass gives   $M_{HI}$ = 6.8 x  10$^{9}$ M$_\odot$ and we use this value for all  calculations requiring the galaxy's  \hi\ mass. \textcolor{black}{ $M_{HI}$ [\msolar] is calculated using  the equation: {\begin{equation}
    M_{HI} = 2.356 \times 10^5 D^2 \int SdV 
\end{equation}
Where, D = distance [Mpc] and $\int S dV$ = flux density [Jy \km].}}

\begin{figure}
\centering
\includegraphics[scale=0.7]{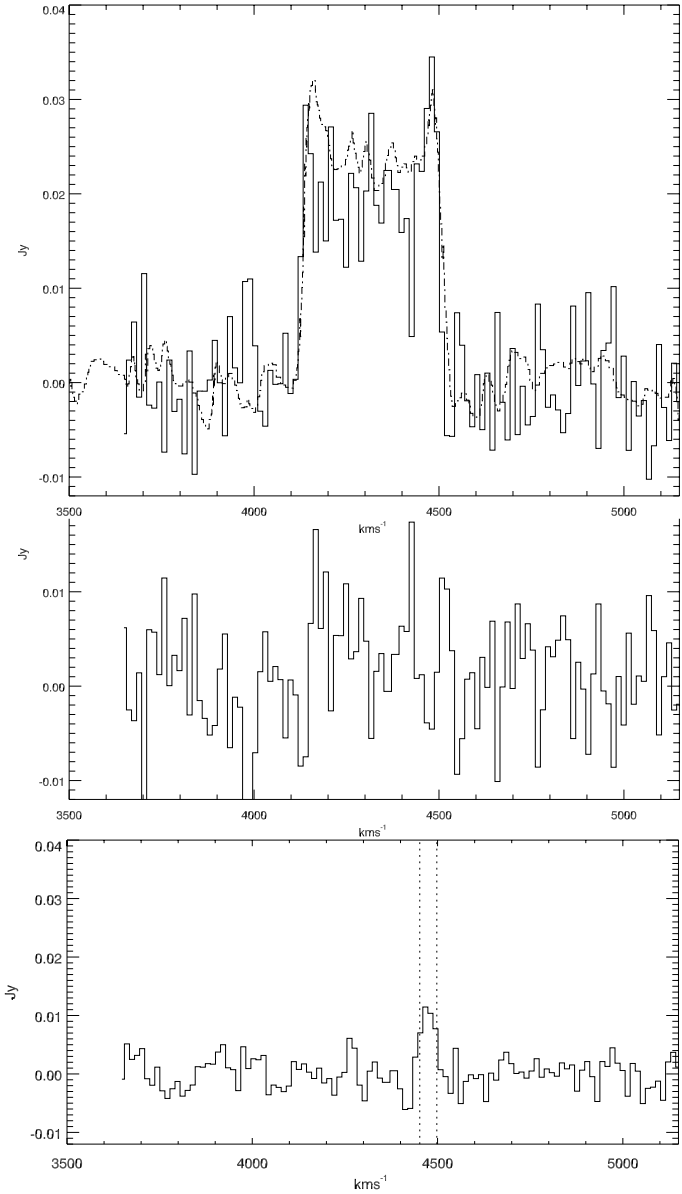}  
\caption{\hi\ spectra for CIG\,340. \textcolor{black}{\textit{\textbf{Above:}}} Single dish from \cite{lewis85}  (dashed line) and from GMRT -- channel width 13.7 \km\ (solid line). \textcolor{black}{ \textcolor{black}{\textit{\textbf{Centre:}}} Single dish minus GMRT residuals.} \textcolor{black}{\textcolor{black}{\textit{\textbf{Below:}}} GMRT \hi\ extension  spectrum, with the dashed vertical lines indicating its the W$_{50}$ velocity width (27 \km). }}
\label{spect}
\end{figure}

 Figure \ref{spect} \textcolor{black}{(top panel)  compares the GMRT  \hi\ }spectrum (solid line) with the \textcolor{black}{manually digitized} single dish spectrum (dashed line) from \cite{lewis85}. The single dish spectrum displays a double horn profile, \textcolor{black}{with the}  \hi\ flux in the low velocity horn ($\sim$ 4150 \km) \textcolor{black}{ being quite similar to that in} the high velocity horn ($\sim$ 4475\km). \textcolor{black}{The \textcolor{black}{middle} panel of Figure \ref{spect} shows \textcolor{black}{that,  within the noise, } the profile of our lower velocity resolution (13.7 \km)  GMRT  spectrum is  in good agreement with the single dish spectrum.} \textcolor{black}{ This conclusion is supported by  the 0.93 ratio between the sum of the  residuals  in the velocity range over which \hi\ was detected,  above ($\geq$ 4336 \km $<$ 4542 \km) and below  ($>$ 4110 \km $<$ 4336 \km) the \hi\ systemic velocity.}

Taking the major axis diameter as $\sim$ 1.84$^{\prime}$ (Table \ref{table_a}) gives a  log$({{\frac{M_{H_{I}}}{D_{l}^{2}}}})$  value\textcolor{black}{ \footnote{$M_{HI}$ = \msolar\ and  $D_{l}$ = $D_{25}$ in kpc}} of 6.89. A comparison with the \hi\ surface densities of  isolated field galaxies of a similar morphological type \citep{hayn84} shows that CIG\,340 has \textcolor{black}{a} normal \hi\ content. The AMIGA sample is more isolated than the  \cite{hayn84} sample so we compared the single dish \hi\ mass  with our AMIGA sample as well. The comparison with the AMIGA sample confirms the \hi\ content of CIG\,340 is  similar to  that of an isolated galaxy of its morphological type and size  (Espada  \textcolor{black}{ private communication}).

The AMIGA isolation parameters \textcolor{black}{ based on either the \citep{verley07b} or  SDSS photometric data}  may not include optically faint gas rich companions. Therefore, the \hi\ cube was searched  for previously undetected companions over the entire  GMRT primary beam, (the L band FWHM is $\sim$ 24$^{\prime}$ which is $\sim$ \textcolor{black}{380} kpc at the distance of CIG\,340),  and \textcolor{black}{ covering a velocity range of $\sim$ 1500 \km.}  Although the signal to noise ratio (SNR) falls considerably beyond the primary beam FWHM, an area twice the FWHM diameter of ($\sim$\textcolor{black}{0.75} Mpc)  was also searched but no \hi\  companions were detected in either search. The \hi\ mass detection limit \textcolor{black}{for the low resolution cube } inside the FWHM of the GMRT primary beam, assuming a  3 channel line width and 3 $\sigma$ SNR, is \textcolor{black}{$\sim$ \textcolor{black}{1.2} x  $\times$10$^{8}$ \msolar.}

\subsection {HI velocity field and rotation}
\label{velo}
\begin{figure}
\centering
\includegraphics[scale=0.48]{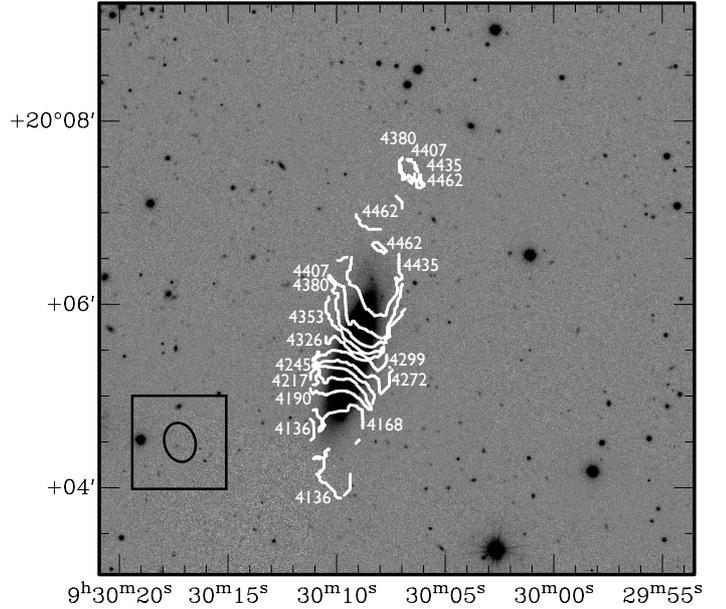}  
\caption{\hi\ velocity field contours \textcolor{black}{in \km}\, derived from the high resolution cube \textcolor{black}{overplotted}  on \textcolor{black}{a} SDSS \textit{r} -- band image. \textcolor{black}{The velocity contours are plotted in steps \textcolor{black}{of $\sim$ 27 \km\ and  the systemic radial velocity is } 4336 \km\  \citep{lewis85}. The spatial resolution is \textcolor{black}{26.08$^{\prime\prime}$  x 19.97$^{\prime\prime}$}  with the beam size  indicated by the ellipse at the bottom left. }}
\label{velfield}
\end{figure}

\begin{figure}
\centering
\includegraphics[scale=0.55]{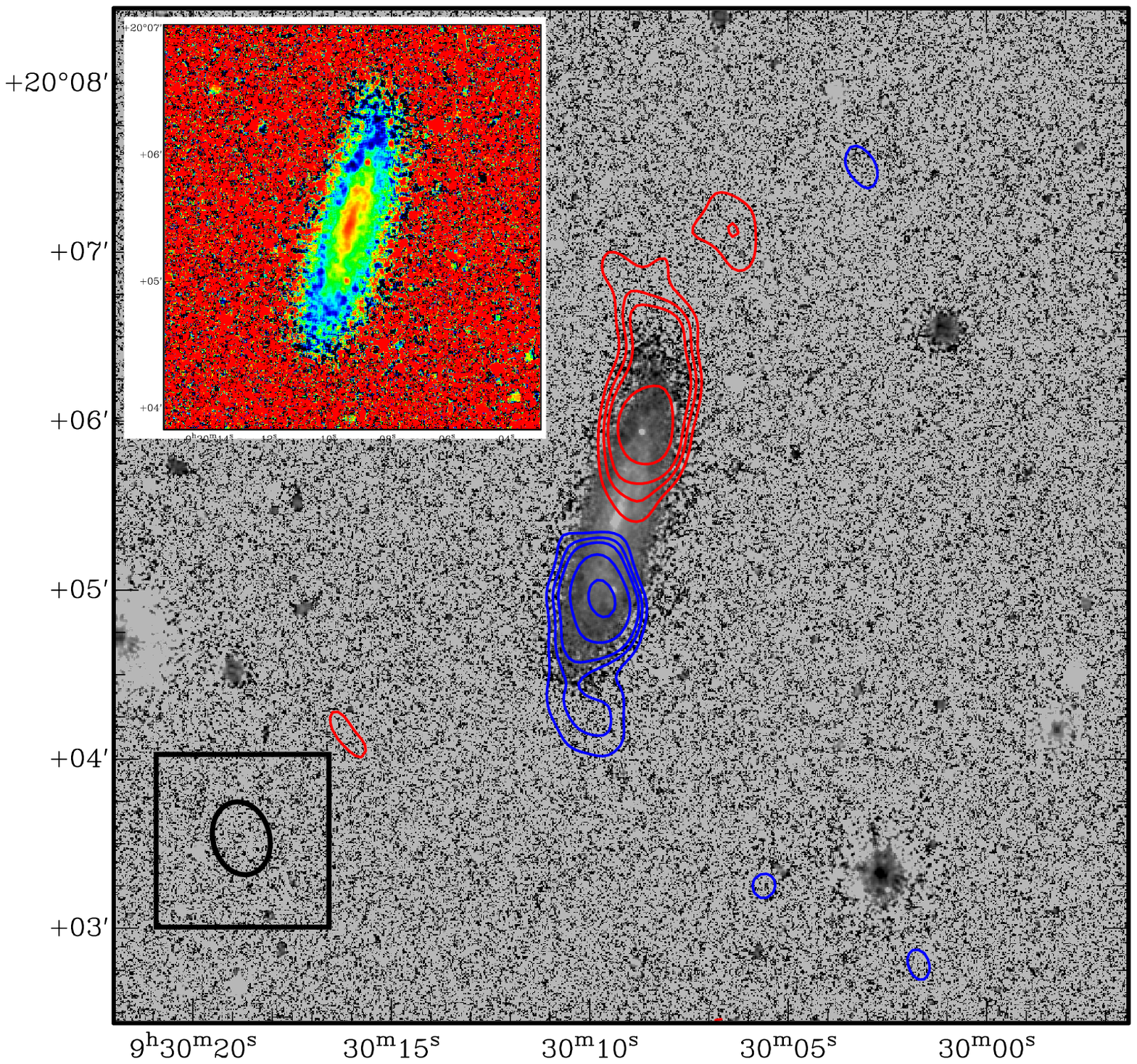}  
\caption{\textcolor{black}{  \hi\ contours for the 4163 \km\ (blue) and 4462 \km\ (red) channels maps from the high resolution cube overlaid on  a greyscale SDSS \textit{g -- i} band image of the galaxy.  The \hi\ contours are  at  3, 4 , 5, 7.5 and 10 $\sigma$ levels with the beam shown at the bottom left as a black ellipse.\textit{ \textbf{Insert:}} zoom in of SDSS \textit{g -- i} band image where the false colour red indicates  g--i $\sim$ 0.25 and blue,  g--i $\sim$0.55}  }
\label{g-i}
\end{figure}

\textcolor{black}{Systemic velocities  derived from \hi\  \citep[4336 \km,][]{lewis85}  and optical  \citep[4339  $\pm$ 6 \km,][]{fernadez12} observations are in good agreement.  \hi\ emission was detected in the high resolution cube at velocities between  4136 \km\ and 4489 \km\ ($\Delta$V =353 \km) and  between 4109 \km\ and 4489 \km\ ($\Delta$V =380 \km) in  the low resolution cube. Figure \ref{velfield} shows the \hi\ velocity field from the high resolution cube. Contours in the velocity field at the northern edge of the disk are unreliable because of low SNR artefacts.   The changing angle of the \hi\ iso--velocity contours in the velocity field, from PA = 104\degree\ at 4168 \km\ to PA = 51\degree at 4435 \km, indicate a warped \hi\ disk \textcolor{black}{\citep{bosma1978}}. \textcolor{black}{The warp is also clearly seen in Figure \ref{g-i} which shows the \hi\ contours from the 4163 \km\ (blue) and 4462 \km\ (red) channels from the high resolution cube overlaid on  an SDSS \textit{g -- i} band image.  } }

\begin{figure}
\centering
\includegraphics[scale=0.8]{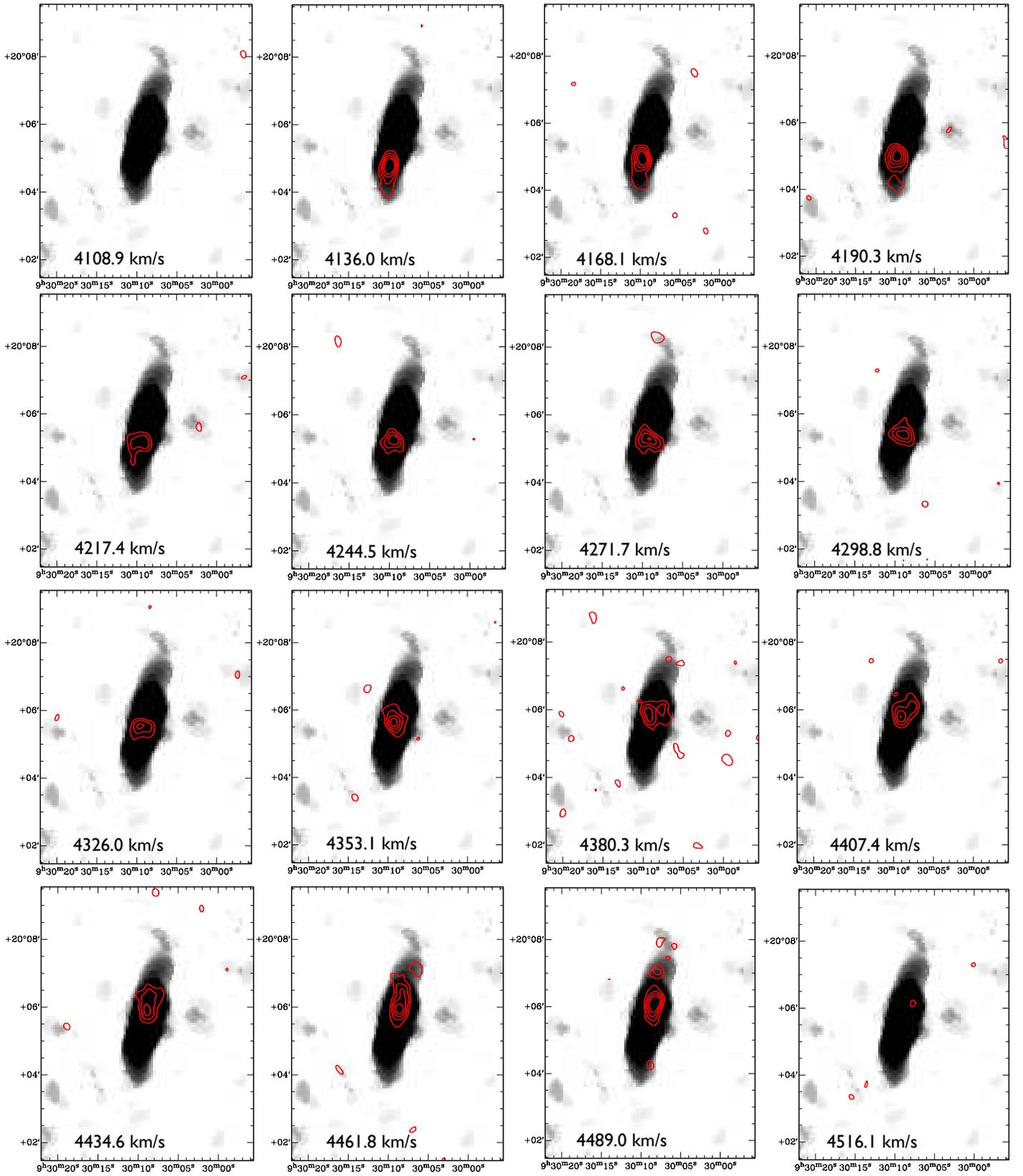}  
\caption{Channel maps:  Contours (red) from the high resolution 26.08 $^{\prime\prime}$ $\times$ 19.97 $^{\prime\prime}$ cube are  overlayed on a  low resolution greyscale integrated H{\sc i} map, with the central velocity of each channel  shown in the bottom left corner of each frame. Contour levels are at 3, 5, 7 and 9 $\sigma$ where 3 $\sigma$  corresponds to a column density of \textcolor{black}{1.7 x  10$^{20}$ cm$^{-2}$.} }
\label{channel_maps}
\end{figure}

 In Figure \ref{channel_maps} we display the channel maps \textcolor{black}{from the high resolution cube which shows } the \hi\ emission at the indicated heliocentric velocities (channel width of $\sim$27  \km) with  rms noise of \textcolor{black}{1.0} mJy/ beam in each channel, implying a \textcolor{black}{3 $\sigma$ detection threshold level of \textcolor{black}{1.7}} x  10$^{20}$ cm$^{-2}$. The \hi\ contours \textcolor{black}{are} overlaid on the \hi\ integrated \textcolor{black}{low resolution  greyscale image}. The \textcolor{black}{spectrum of the 12 kpc \hi\ extension, referred to in the previous section, is shown in the Figure \ref{spect} -- lower panel and is also } visible in the 4462 \km\ and  4489 \km\ channel maps. \textcolor{black}{ Evidence  of the warped \hi\ disk  \textcolor{black}{also} comes from comparing the CIG\,340 channel maps to those from  warp models in  \cite{swaters97,gentile03}.  In those warp models  several intermediate channel maps  between the systemic and extreme velocities  display   \textcolor{black}{ asymmetric bifurcations}, approximately aligned with the galaxy major axis. In CIG\,340 we see indications of  \textcolor{black}{ asymmetric bifurcations} in the 4244.5 \km, 4271.7\km, 4380.3 \km, and 4407.4 \km\ channel maps.  The \aflux\ parameter derived from the GMRT spectrum was  = 1.01 and deducting   \hi\ extension spectrum from the GMRT \textcolor{black}{spectrum} increased the  \aflux\ parameter to 1.10 \textcolor{black}{($>$  3 $\sigma$ from Table \ref{table_a})}. \textcolor{black}{This increase in kinematic asymmetry, on removal of the  \hi\ extension spectrum,  suggests that the \hi\ extension is a morphological distortion within a kinematically symmetric \hi\ disk. }}    

\subsection {\textcolor{black}{Position -- Velocity (PV) diagrams}}
\label{pvdiag_sec}
\begin{figure*}
\centering
\includegraphics[scale=0.8]{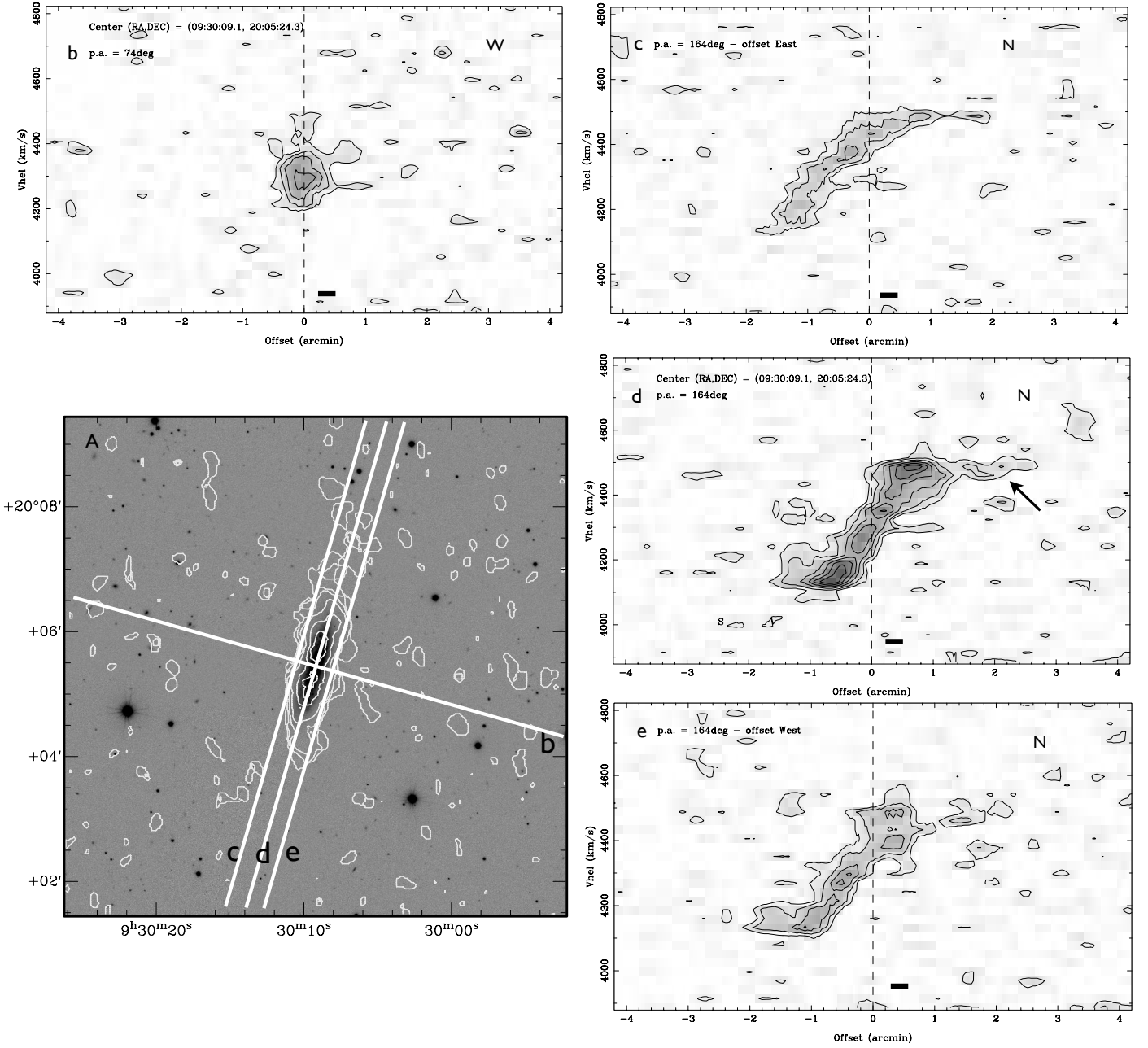}  
\caption {The orientation of each  PV diagram cut from the high resolution cube, shown in panels  b to e,  is \textcolor{black}{indicated in panel A  against a background  SDSS \textit{r } -- band image with \hi\ contours from the high resolution cube.  The vertical axis in each  PV diagram  gives the velocity in \km\ and  the horizontal axis gives  the offset in arc seconds as follows:   diagram \textbf{ b)} along an axis perpendicular , PA = 74\degree, to the major axis centred at the optical centre, 09:30:9.07 + 20:05:25,  \textbf{c)} offset 20 arcsec east and parallel to  the optical major axis PA = 164\degree , \textbf{d)} along the  optical major axis, PA = 164\degree,  centred at the optical centre 09:30:8.62 +20:05:56. \textbf{e)} offset 20 arcsec west and parallel to  the optical major axis PA = 164\degree.  For diagrams \textbf{c} to \textbf{e} positive offset values are north of the optical centre. \textcolor{black}{In diagram  \textbf{c}) the arrow indicates the \hi\ extension.} The  \hi\ systemic velocity of the galaxy is 4336 \km\ \citep{lewis85}.  The black bars indicate the approximate size of the high resolution beam.  }}
\label{pvdiag}
\end{figure*}

\textcolor{black}{  \hi\  PV diagrams from the high resolution cube for cuts along the major (PA =164\degree) and minor (PA = 74\degree) axes centred at the galaxy's optical centre are shown in Figure \ref{pvdiag} -- panels d and b.  Additionally the Figure shows cuts parallel to the major axis cut, 20 arcsec to the east (panel c) and 20 arcsec to the west (panel e). } Overall the major axis PV diagram is consistent with a rotating  edge--on \hi\ disk, although  there are  differences in velocity structure north and south of the optical center. \textcolor{black}{For example north of the optical center the velocity range  between offsets from the optical center  of +1.3 arcmin and +2.7 arcmin is $\sim$ 50 \km. This contrasts to the south where the corresponding value between offsets at -1.3 and -1.7 arcsec  is $\sim$100 \km.} Whatever the mechanism producing the larger \hi\ \textcolor{black}{velocity} range in south it could also have dispersed the \hi\ there to column densities below the GMRT detection threshold, but, within the noise, the residuals in Figure  \ref{spect}   do not confirm this. \textcolor{black}{The \hi\ extension, marked with an arrow in Figure \ref{pvdiag} -- d, is kinematically clearly a continuation of the rotation  curve.  }

For two small edge--on spirals  \cite{kamphuis13} were able to use PV diagrams perpendicular to the major axis from deep \hi\ observations of two edge--on spirals  together with  models to differentiate between a warped disk with spiral arms and an  un--warped disk containing asymmetric \hi\ distributions. Unfortunately our CIG\,340 observation lacks the sensitivity and velocity resolution  to detect the faint features required to make the comparison with models, \textcolor{black}{ although Figure \ref{pvdiag} -- b  shows  \hi\ is detected up to 1.2 arcmin (19 kpc) above (west) of the optical centre. The} lack of symmetry between the PV diagrams for the cuts parallel to  major axis provides further evidence of the warped \hi\ disk. 

\subsection{ \textcolor{black}{\hi\ and stellar  morphologies }}
\label{morph}
 
\textcolor{black}{ The warp in the stellar disk  is quite evident at the northern edge of the optical disk (Figure \ref{sdss} -- right panel). \textcolor{black}{Warping at both northern and southern edges of the UV disk is also seen in the  \textit{GALEX} FUV and NUV images (not shown). } But the warp is  much more apparent and extensive in \hi\ (Figures \ref{mom0} and \ref{g-i}) and it seems likely that at least  a galactic rotation ($\sim$ 0.5 Gyr) \textcolor{black}{would be} required for the warp to become a disk wide phenomenon.  }
 
\textcolor{black}{A bright   lens is visible at the galaxy centre  in both the SDSS  (Figure  \ref{sdss}) and 2MASS images, however  lenses  are unusual in a galaxy with  a morphological type as late as Sbc.  There is no indication from the optical images that a central classical bulge, primarily consisting of old stellar populations, extends above the optical disk, This is consistent with the absence of clear evidence for a  central \hi\ hole, however we do see a depression in the \hi\ at the optical centre. } 
 
\begin{figure}
\centering
\includegraphics[scale=0.37]{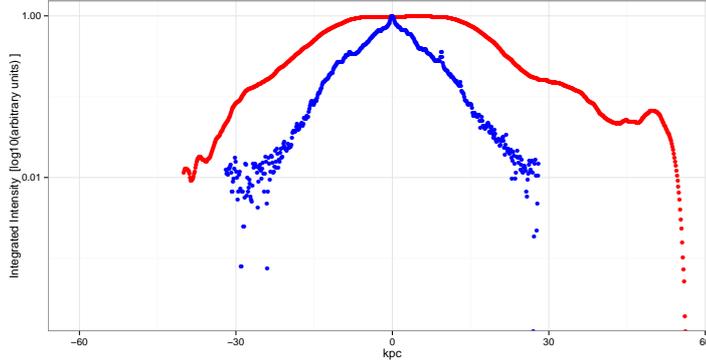}  
\caption{Integrated \hi\  flux \textcolor{black}{(log10 scale)} from the low resolution map (red) is compared to the integrated SDSS \textit{ i} -- band emission along the major axis (blue).  The kpc scale show offset from the optical centre along the major axis, with negative values to south and positive values to the north of the optical centre. The \textit{i} -- band and \hi\ emission have been normalised to make the distributions comparable.  }
\label{hi_maj}
\end{figure}

\textcolor{black}{To further investigate the symmetry of \hi\ and stellar disks we  integrated the  SDSS \textit{i} -- band emission (as a proxy for the older stellar population) and  \hi\ emission  along the galaxy's  major axis (PA = \textcolor{black}{164}\degree). This analysis \textcolor{black}{confirms} the stellar emission is quite symmetrically  distributed north and south of the optical centre (north/south emission ratio =0.95). In contrast the \hi\ north/south emission ratio is 1.32.   Figure \ref{hi_maj} shows  the  log of the normalised  integrated \hi\ emission (red) compared to the normalised integrated SDSS \textit{ i} -- band emission (blue) along the optical major axis, south (left) and north (right) of the optical centre. From Figure \ref{hi_maj} the difference between the symmetrically distributed \textit{i } -- band  light  and the lopsided  \hi\ distribution  favouring  the north  is clearly apparent.  Moreover the \hi\  in the figure shows a clear departure from an exponential decline with distance from the optical centre, starting $\sim$ 106\prin\ (28 kpc) north of the  optical centre. This departure from the general distribution of \hi\  in the disk suggests  that the \textcolor{black}{\hi\ extension} could be up to $\sim$64\prin\ (17 kpc) in length.}

\section{Discussion}
\label{discuss}

\textcolor{black}{First we  discuss the isolation of CIG\,340 then briefly consider several possible causes of the  observed \hi\ asymmetries focusing primarily on the most unexpected feature, the \hi\ extension}.

\textcolor{black}{\textit{\textbf{Isolation}}} \textcolor{black}{  CIG 340 (mag=14.1g, 4339 \km) is part of the low--density-contrast group, LDCE 0651, identified in the  Two Micron All-Sky Redshift Survey by \cite{crook08}. The other group members  are CGCG\,91--099 (mag = 15.3, 4362 \km) and CGCG\,92--005 (mag = 14.8g, 4456 \km).  Of these neighbours, CGCG\,91--099 is  projected   nearest  to CIG\,340, at a distance of  37.438\prim\ (0.6 Mpc). To cover this projected distance at a velocity typical of group members (200 \km) would take $\sim$ 2.9 Gyrs,  which is 6 times longer than the 0.5 Gyr for a single galaxy rotation\footnote{\textcolor{black}{T$_{rot}$ [Gyr]= 6.1478 r/ V$_{rot}$, where r = the optical radius [kpc] and V$_{rot}$= 0.5 $\Delta$V  [\km]/ sin(i).}},  based on $\Delta$V$_{HI}$  = 395.9 \km\ \citep{lewis85}.  It is generally considered that both stellar and \hi\ asymmetries arising from tidal  interactions will dissipate within a single galactic rotation.   More  particularly   \hi\ asymmetries caused by major merger interactions remain detectable for only $\sim$ 0.4 Gyr to 0.7 Gyr \citep{holwerda11}, i.e., approximately   a  single  rotation period for  CIG\,340. Applying the density and tidal force parameters from \cite{argudo13} to  neighbouring galaxies\footnote {which includes both LDCE 0651 neighbours} within a  1 Mpc radius, with optical diameters between 0.25 and 4 times that of the CIG\,340, we conclude there have been no major interactions  with currently observed major companions  within at least the last 5.2 Gyr. So a recent major interaction with a known neighbour of similar size can safely  be ruled out as the cause of any of the stellar or \hi\ asymmetries observed in CIG\,340. }

  While  \hi\ lopsidedness and warps in late--type galaxies are common and have been observed in other isolated galaxies,  the  $\sim$ 12 kpc \textcolor{black}{\hi\ extension}  is unexpected in an isolated galaxy. Below we summarise several possible causes of the \textcolor{black}{\hi\ extension}:


\textit{\textbf{A minor companion which is now disrupted or projected within the disk. }}There are no  indications from the 2MASS \textit{J,H,K} -- band images of a minor companion. The northern outer disk perturbation  in the SDSS \textit{g, r, i} -- band image (Figure \ref{sdss} -- right panel), the \textcolor{black}{\hi\ extension} in Figure \ref{mom0}, and the \textcolor{black}{ larger  \hi\ velocity range (100 \km) at the} southern edge of the \hi\ disk (Figure \ref{pvdiag}) could all conceivably be the impacts of a minor companion on an orbit which has passed through both the  northern  and southern edges of the \hi\ disk. \textcolor{black}{An interesting feature in the SDSS \textit{g-i} image (Figure \ref{g-i} -- insert)  is the presence of a ridge of unusually blue emission running $\sim$ 30 arcsec (7.5 kpc) from in the north of the optical disk  which could be related to passage of such a companion.}

Modelling by \cite{weinberg98} and \cite{weinberg06} showed that a galaxy of the mass of LMC on a polar orbit of a Milky Way mass galaxy can  warp the principal galaxy's stellar and \hi\ disks  to produce an \textcolor{black}{``integral''}  sign warp morphology via a halo perturbation. The stellar mass of CIG\,340 is 2.29 x \textcolor{black}{10$^{10}$} \msolar\  \citep{fernadez13}  $\sim$ \textcolor{black}{50}\% of the Milky Way's stellar mass. Provided the warping  effects of the satellite scale approximately with stellar mass then a CIG\,340 satellite of mass  $\sim$ \textcolor{black}{7.5} x 10$^8$ \msolar\ (apparent magnitude of $\sim$ 17) could have created the observed warp.  Such a satellite should be easily detectable in SDSS images, but none \textcolor{black}{is seen. However} CIG\,340 shares a \textcolor{black}{number of} characteristics with the edge on spiral NGC\,4013 (V$_{radial}$ = 831 \km) including \textcolor{black}{a similar optical size,} a warped \hi\ disk and a one sided \textcolor{black}{\hi\ extension}, together with the absence  of any observable satellite. Deep optical observations ($\mu_R$ = 27.0 mag arcsec$^2$) by \cite{delgardo09} revealed that NGC\,4013 has an extensive system of faint tidal streams which were interpreted as debris from multiple orbits of a now disrupted dwarf companion. From the modelling of the NGC\,4013 system it was concluded  that its \hi\ morphology asymmetries were most likely caused by an interaction with the companion. 

\textit{\textbf{Accretion of a cold \hi\ cloud.}}  The \hi\ mass \textcolor{black}{$\sim$ 4 \textcolor{black}{x} 10$^8$ \msolar.} and diameter $\sim$12 kpc  of the \textcolor{black}{\hi\ extension} are orders of magnitude  greater than those of high velocity and intermediate velocity clouds in the local group  which have  \hi\  masses of $\sim$10$^5$ \msolar\  to 10$^6$ \msolar\  and \hi\ diameters of $\sim$1--3 kpc \citep{wakker99,westmeier08}. \textcolor{black}{ Moreover an \textcolor{black}{infalling} cloud would not be expected to fall on an extension of the rotation curve (see Figure \ref{pvdiag} -- d ).  }\textcolor{black}{This and the comparison  of the \aflux, including and excluding  the \hi\ extension, in section \ref{velo}  provides strong evidence that the \hi\  extension was originally part the outer \hi\ disk rather than recently accreted material.} 

\textit{\textbf{Disk/halo misalignment. }}  In this scenario, which is proposed as an explanation for the frequently observed lopsided \hi\ distributions in late--type galaxies, the centres of the baryonic  disks are misaligned with  the dark matter potential. Modelling indicates that this misalignment \textcolor{black}{would}  dissipate within 5 rotations \citep{dubinski95}, i.e. $\sim$ 2.5 Gyr for CIG\,340.  Moreover high amplitude halo asymmetries from mergers or major interactions are predicted to be detectable for up to 2  Gyr \citep{bournaud05}. \textcolor{black}{So while the  isolation parameters appear to rule out  interactions with the currently observed major companions as the source of the  \textcolor{black}{\hi\ extension} on the basis of time scales,  an asymmetry in the dark matter halo arising from a merger within the last 2 Gyr  could explain both the lopsidedness of the \hi\  distribution  and  lack of a kinematic signature for the \textcolor{black}{extension} as well as the \textcolor{black}{\hi\ } warp. Moreover we cannot discount a halo perturbation originating from an  interaction with a now disrupted satellite \citep{jog09}. }

\textcolor{black}{ In common with \textcolor{black}{other} AMIGA isolated spirals  previously mapped in \hi\  (CIG\,85, CIG\,96 and CIG\,292, section \ref{intro}), CIG\,340 has  indications that at least some its \hi\ asymmetries were generated comparatively recently ($\sim$ 10$^8$ yr) and that those asymmetries were not produced by interactions with \textcolor{black}{the currently observed} major companions.  Like the other three galaxies, the signatures of \hi\ asymmetry are  stronger in the resolved morphology than in the velocity field. \textcolor{black}{The \citep{verley07b} isolation parameters for CIG\,85, CIG\,96 and CIG\,340 indicate these galaxies are well isolated even within the AMIGA sample and signatures  of lopsidedness and warps  are also present in CIG\,85 and CIG\,292.}}
 
 \textcolor{black}{ We  note the} continuum flux reported from NVSS\footnote{NRAO VLA Sky Survey}  is 2.2 mJy  implies a  star formation rate of 0.46  \msolar\ yr$^{-1}$, although continuum flux was not detected with the GMRT,  \textcolor{black}{consistent with a quiescence recent SF history. The radio -- FIR correlation relation \citep{sabater2008} parameters for CIG\,340 are, log L($_{1.4}$) = 20.9  and log(L$_{60}$)  = 9.19,    confirming that the galaxy  lies on the radio -- FIR correlation. }


\section{Concluding remarks}
\label{conclusions}
\textcolor{black}{Unexpectedly} our GMRT \hi\ observations  revealed significant morphological  asymmetries despite CIG\,340 having  a highly symmetric \hi\ \textcolor{black}{spectrum}  (\aflux\ = 1.03 $\pm$ 0.02). The most notable \hi\ features \textcolor{black}{are:} the warped  disk (\textcolor{black}{ with an optical counterpart}), the ratio of \hi\ emission north and south of the optical centre (1.32), and a $\sim$ 45\prin\   (12 kpc) \textcolor{black}{\hi\ extension} \textcolor{black}{at the northern edge of the disk. While the \textcolor{black}{ \hi\ extension} is morphologically distinguishable from main body of the \textcolor{black}{ \hi\ disk suggesting}  a recent perturbation \textcolor{black}{($\sim$10$^8$)}, there is no unambiguous equivalent kinematic signature. Based on \textcolor{black}{its distinct morphology the most likely source of the \textcolor{black}{\hi\ extension} appears to be   an interaction with small currently undetected satellite.\textcolor{black}{ This satellite may also be responsible for producing the lopsidedness and  the warped stellar / \hi\ disks.} Although we cannot rule out the possibility that \textcolor{black}{the \hi\ tail lopsidedness and  the warped stellar / \hi\ disks are evidence of  a  halo asymmetry produced by } a merger that occurred between $\sim$0.5 Gyr and 2 Gyr ago, although we have no direct evidence for such a merger}.  CIG\,340 demonstrates that in isolated galaxies a highly symmetric \hi\ spectrum can still mask significant \hi\ morphological asymmetries which can be   revealed by \hi\ interferometric mapping.}

 \textcolor{black}{Deep optical  observations to determine the presence or absence of stellar trails would help constrain the minor companion and cold gas cloud accretion scenarios. These observations  together with a  high resolution kinematic study of the \textcolor{black}{central parts of the galaxy}  could provide \textcolor{black}{ further} insights into  the origin of the  \hi\ asymmetries, including the warp.}

\section {Acknowledgements}
\textcolor{black}{We are grateful to the anonymous referee for 
his/her helpful remarks, which have improved the presentation of this paper.} This work has been supported by Grant AYA2011-30491-C02-01 co-financed by MICINN and FEDER funds, and the Junta de Andalucia (Spain) grants P08-FQM-4205 and TIC-114. We  \textcolor{black}{also thank } the AMIGA team for their comments and suggestions.
We thank the staff of the GMRT who have made these observations
  possible. GMRT is run by the National Centre for Radio Astrophysics
  of the Tata Institute of Fundamental Research.
The Nasa Extragalactic Database, NED, is operated by the Jet Propulsion Laboratory, California Institute of Technology, under contract with the National Aeronautics and Space Administration.
We acknowledge the usage of the HyperLeda database (http://leda.univ-lyon1.fr).
Funding for the SDSS and SDSS-II has been provided by the Alfred P. Sloan Foundation, the Participating Institutions, the National Science Foundation, the U.S. Department of Energy, the National Aeronautics and Space Administration, the Japanese Monbukagakusho, the Max Planck Society, and the Higher Education Funding Council for England. The SDSS Web Site is http://www.sdss.org/. The SDSS is managed by the Astrophysical Research Consortium for the Participating Institutions. The Participating 
Institutions are the American Museum of Natural History, Astrophysical Institute Potsdam, University of Basel, University of Cambridge, Case Western Reserve University, University of Chicago, Drexel University, Fermilab, the Institute for Advanced Study, the Japan Participation Group, Johns Hopkins University, the Joint Institute for Nuclear Astrophysics, the Kavli Institute for Particle Astrophysics and Cosmology, the Korean Scientist Group, the Chinese Academy of Sciences (LAMOST), Los Alamos National Laboratory, the Max-Planck-Institute for Astronomy (MPIA), the Max-Planck-Institute for Astrophysics (MPA), New Mexico State University, Ohio State University, University of Pittsburgh, University of Portsmouth, Princeton University, the United States Naval Observatory, and the University of Washington.  Based on observations made with the NASA Galaxy

\bibliographystyle{aa} 
\bibliography{cig-v8}

\begin{thebibliography}{41}
\expandafter\ifx\csname natexlab\endcsname\relax\def\natexlab#1{#1}\fi

\bibitem[{{Argudo-Fern{\'a}ndez} {et~al.}(2013){Argudo-Fern{\'a}ndez},
  {Verley}, {Bergond}, {Sulentic}, {Sabater}, {Fern{\'a}ndez Lorenzo}, {Leon},
  {Espada}, {Verdes-Montenegro}, {Santander-Vela}, {Ruiz}, \&
  {S{\'a}nchez-Exp{\'o}sito}}]{argudo13}
{Argudo-Fern{\'a}ndez}, M., {Verley}, S., {Bergond}, G., {et~al.} 2013, A\&A,
  560, A9

\bibitem[{{Baars} {et~al.}(1977){Baars}, {Genzel}, {Pauliny-Toth}, \&
  {Witzel}}]{baars77}
{Baars}, J.~W.~M., {Genzel}, R., {Pauliny-Toth}, I.~I.~K., \& {Witzel}, A.
  1977, A\&A, 61, 99

\bibitem[{{Bergvall} \& {Ronnback}(1995)}]{bergvall95}
{Bergvall}, N. \& {Ronnback}, J. 1995, MNRAS, 273, 603

\bibitem[{{Bosma}(1978)}]{bosma1978}
{Bosma}, A. 1978, PhD thesis, PhD Thesis, Groningen Univ., (1978)

\bibitem[{{Bournaud} \& {Combes}(2002)}]{bournaud02}
{Bournaud}, F. \& {Combes}, F. 2002, A\&A, 392, 83

\bibitem[{{Bournaud} {et~al.}(2005){Bournaud}, {Combes}, {Jog}, \&
  {Puerari}}]{bournaud05}
{Bournaud}, F., {Combes}, F., {Jog}, C.~J., \& {Puerari}, I. 2005, A\&A, 438,
  507

\bibitem[{{Crook} {et~al.}(2008){Crook}, {Huchra}, {Martimbeau}, {Masters},
  {Jarrett}, \& {Macri}}]{crook08}
{Crook}, A.~C., {Huchra}, J.~P., {Martimbeau}, N., {et~al.} 2008, ApJ, 685,
  1320

\bibitem[{{Dubinski} \& {Kuijken}(1995)}]{dubinski95}
{Dubinski}, J. \& {Kuijken}, K. 1995, ApJ, 442, 492

\bibitem[{{Durbala} {et~al.}(2008){Durbala}, {Sulentic}, {Buta}, \&
  {Verdes-Montenegro}}]{durbala08}
{Durbala}, A., {Sulentic}, J.~W., {Buta}, R., \& {Verdes-Montenegro}, L. 2008,
  \mnras, 390, 881

\bibitem[{{Espada} {et~al.}(2005){Espada}, {Bosma}, {Verdes-Montenegro},
  {Athanassoula}, {Leon}, {Sulentic}, \& {Yun}}]{espada05}
{Espada}, D., {Bosma}, A., {Verdes-Montenegro}, L., {et~al.} 2005, \aap, 442,
  455

\bibitem[{{Espada} {et~al.}(2011{\natexlab{a}}){Espada}, {Mu{\~n}oz-Mateos},
  {Gil de Paz}, {Sabater}, {Boissier}, {Verley}, {Athanassoula}, {Bosma},
  {Leon}, {Verdes-Montenegro}, {Yun}, \& {Sulentic}}]{espada11b}
{Espada}, D., {Mu{\~n}oz-Mateos}, J.~C., {Gil de Paz}, A., {et~al.}
  2011{\natexlab{a}}, ApJ, 736, 20

\bibitem[{{Espada} {et~al.}(2011{\natexlab{b}}){Espada}, {Verdes-Montenegro},
  {Huchtmeier}, {Sulentic}, {Verley}, {Leon}, \& {Sabater}}]{espada11}
{Espada}, D., {Verdes-Montenegro}, L., {Huchtmeier}, W.~K., {et~al.}
  2011{\natexlab{b}}, A\&A, 532, A117

\bibitem[{{Fern{\'a}ndez Lorenzo} {et~al.}(2013){Fern{\'a}ndez Lorenzo},
  {Sulentic}, {Verdes-Montenegro}, \& {Argudo-Fern{\'a}ndez}}]{fernadez13}
{Fern{\'a}ndez Lorenzo}, M., {Sulentic}, J., {Verdes-Montenegro}, L., \&
  {Argudo-Fern{\'a}ndez}, M. 2013, MNRAS, 434, 325

\bibitem[{{Fern{\'a}ndez Lorenzo} {et~al.}(2012){Fern{\'a}ndez Lorenzo},
  {Sulentic}, {Verdes-Montenegro}, {Ruiz}, {Sabater}, \&
  {S{\'a}nchez}}]{fernadez12}
{Fern{\'a}ndez Lorenzo}, M., {Sulentic}, J., {Verdes-Montenegro}, L., {et~al.}
  2012, ArXiv e-prints

\bibitem[{{Gentile} {et~al.}(2003){Gentile}, {Fraternali}, {Klein}, \&
  {Salucci}}]{gentile03}
{Gentile}, G., {Fraternali}, F., {Klein}, U., \& {Salucci}, P. 2003, A\&A, 405,
  969

\bibitem[{{Haynes} \& {Giovanelli}(1984)}]{hayn84}
{Haynes}, M.~P. \& {Giovanelli}, R. 1984, AJ, 89, 758

\bibitem[{{Hern{\'a}ndez-Toledo} {et~al.}(2008){Hern{\'a}ndez-Toledo},
  {V{\'a}zquez-Mata}, {Mart{\'{\i}}nez-V{\'a}zquez}, {Avila Reese},
  {M{\'e}ndez-Hern{\'a}ndez}, {Ortega-Esbr{\'{\i}}}, \&
  {N{\'u}{\~n}ez}}]{hernadez08}
{Hern{\'a}ndez-Toledo}, H.~M., {V{\'a}zquez-Mata}, J.~A.,
  {Mart{\'{\i}}nez-V{\'a}zquez}, L.~A., {et~al.} 2008, AJ, 136, 2115

\bibitem[{{Holwerda} {et~al.}(2011){Holwerda}, {Pirzkal}, {Cox}, {de Blok},
  {Weniger}, {Bouchard}, {Blyth}, \& {van der Heyden}}]{holwerda11}
{Holwerda}, B.~W., {Pirzkal}, N., {Cox}, T.~J., {et~al.} 2011, MNRAS, 416, 2426

\bibitem[{{Jog} \& {Combes}(2009)}]{jog09}
{Jog}, C.~J. \& {Combes}, F. 2009, Physics Reports, 471, 75

\bibitem[{{Kamphuis} {et~al.}(2013){Kamphuis}, {Rand}, {J{\'o}zsa},
  {Zschaechner}, {Heald}, {Patterson}, {Gentile}, {Walterbos}, {Serra}, \& {de
  Blok}}]{kamphuis13}
{Kamphuis}, P., {Rand}, R.~J., {J{\'o}zsa}, G.~I.~G., {et~al.} 2013, MNRAS,
  434, 2069

\bibitem[{{Leon} \& {Verdes-Montenegro}(2003)}]{leon03}
{Leon}, S. \& {Verdes-Montenegro}, L. 2003, A\&A, 411, 391

\bibitem[{{Leon} {et~al.}(2008){Leon}, {Verdes-Montenegro}, {Sabater},
  {Espada}, {Lisenfeld}, {Ballu}, {Sulentic}, {Verley}, {Bergond}, \&
  {Garc{\'{\i}}a}}]{leon2008}
{Leon}, S., {Verdes-Montenegro}, L., {Sabater}, J., {et~al.} 2008, \aap, 485,
  475

\bibitem[{{Lewis} {et~al.}(1985){Lewis}, {Helou}, \& {Salpeter}}]{lewis85}
{Lewis}, B.~M., {Helou}, G., \& {Salpeter}, E.~E. 1985, ApJS, 59, 161

\bibitem[{{Lisenfeld} {et~al.}(2011){Lisenfeld}, {Espada}, {Verdes-Montenegro},
  {Kuno}, {Leon}, {Sabater}, {Sato}, {Sulentic}, {Verley}, \&
  {Yun}}]{lisenfeld2011}
{Lisenfeld}, U., {Espada}, D., {Verdes-Montenegro}, L., {et~al.} 2011, \aap,
  534, A102

\bibitem[{{Lisenfeld} {et~al.}(2007){Lisenfeld}, {Verdes-Montenegro},
  {Sulentic}, {Leon}, {Espada}, {Bergond}, {Garc{\'{\i}}a}, {Sabater},
  {Santander-Vela}, \& {Verley}}]{lisen07}
{Lisenfeld}, U., {Verdes-Montenegro}, L., {Sulentic}, J., {et~al.} 2007, A\&A,
  462, 507

\bibitem[{{Mart{\'{\i}}nez-Delgado} {et~al.}(2009){Mart{\'{\i}}nez-Delgado},
  {Pohlen}, {Gabany}, {Majewski}, {Pe{\~n}arrubia}, \& {Palma}}]{delgardo09}
{Mart{\'{\i}}nez-Delgado}, D., {Pohlen}, M., {Gabany}, R.~J., {et~al.} 2009,
  ApJ, 692, 955

\bibitem[{{Pisano} {et~al.}(2002){Pisano}, {Wilcots}, \& {Liu}}]{pisano02}
{Pisano}, D.~J., {Wilcots}, E.~M., \& {Liu}, C.~T. 2002, ApJS, 142, 161

\bibitem[{{Portas} {et~al.}(2011){Portas}, {Scott}, {Brinks}, {Bosma},
  {Verdes-Montenegro}, {Heesen}, {Espada}, {Verley}, {Sulentic}, {Sengupta},
  {Athanassoula}, \& {Yun}}]{portas}
{Portas}, A., {Scott}, T.~C., {Brinks}, E., {et~al.} 2011, \apjl, 739, L27

\bibitem[{{Sabater} {et~al.}(2008){Sabater}, {Leon}, {Verdes-Montenegro},
  {Lisenfeld}, {Sulentic}, \& {Verley}}]{sabater2008}
{Sabater}, J., {Leon}, S., {Verdes-Montenegro}, L., {et~al.} 2008, \aap, 486,
  73

\bibitem[{{Sancisi} {et~al.}(2008){Sancisi}, {Fraternali}, {Oosterloo}, \& {van
  der Hulst}}]{sancisi08}
{Sancisi}, R., {Fraternali}, F., {Oosterloo}, T., \& {van der Hulst}, T. 2008,
  A\&AR, 15, 189

\bibitem[{{Sengupta} {et~al.}(2012){Sengupta}, {Scott}, {Verdes Montenegro},
  {Bosma}, {Verley}, {Vilchez}, {Durbala}, {Fern{\'a}ndez Lorenzo}, {Espada},
  {Yun}, {Athanassoula}, {Sulentic}, \& {Portas}}]{sengupta12}
{Sengupta}, C., {Scott}, T.~C., {Verdes Montenegro}, L., {et~al.} 2012, A\&A,
  546, A95

\bibitem[{{Struck}(1999)}]{struck99}
{Struck}, C. 1999, Physics Reports, 321, 1

\bibitem[{{Swaters} {et~al.}(1997){Swaters}, {Sancisi}, \& {van der
  Hulst}}]{swaters97}
{Swaters}, R.~A., {Sancisi}, R., \& {van der Hulst}, J.~M. 1997, ApJ, 491, 140

\bibitem[{{Verdes-Montenegro} {et~al.}(2005){Verdes-Montenegro}, {Sulentic},
  {Lisenfeld}, {Leon}, {Espada}, {Garcia}, {Sabater}, \& {Verley}}]{vm05}
{Verdes-Montenegro}, L., {Sulentic}, J., {Lisenfeld}, U., {et~al.} 2005, A\&A,
  436, 443

\bibitem[{{Verley} {et~al.}(2007b){Verley}, {Leon}, {Verdes-Montenegro},
  {Combes}, {Sabater}, {Sulentic}, {Bergond}, {Espada}, {Garc{\'{\i}}a},
  {Lisenfeld}, \& {Odewahn}}]{verley07b}
{Verley}, S., {Leon}, S., {Verdes-Montenegro}, L., {et~al.} 2007b, A\&A, 472,
  121

\bibitem[{{Verley} {et~al.}(2007a){Verley}, {Odewahn}, {Verdes-Montenegro},
  {Leon}, {Combes}, {Sulentic}, {Bergond}, {Espada}, {Garc{\'{\i}}a},
  {Lisenfeld}, \& {Sabater}}]{verley07a}
{Verley}, S., {Odewahn}, S.~C., {Verdes-Montenegro}, L., {et~al.} 2007a, \aap,
  470, 505

\bibitem[{{Vollmer} {et~al.}(2008){Vollmer}, {Braine}, {Pappalardo}, \&
  {Hily-Blant}}]{voll08}
{Vollmer}, B., {Braine}, J., {Pappalardo}, C., \& {Hily-Blant}, P. 2008, A\&A,
  491, 455

\bibitem[{{Wakker} {et~al.}(1999){Wakker}, {Howk}, {Savage}, {van Woerden},
  {Tufte}, {Schwarz}, {Benjamin}, {Reynolds}, {Peletier}, \&
  {Kalberla}}]{wakker99}
{Wakker}, B.~P., {Howk}, J.~C., {Savage}, B.~D., {et~al.} 1999, Nature, 402,
  388

\bibitem[{{Weinberg}(1998)}]{weinberg98}
{Weinberg}, M.~D. 1998, MNRAS, 299, 499

\bibitem[{{Weinberg} \& {Blitz}(2006)}]{weinberg06}
{Weinberg}, M.~D. \& {Blitz}, L. 2006, ApJL, 641, L33

\bibitem[{{Westmeier} {et~al.}(2008){Westmeier}, {Br{\"u}ns}, \&
  {Kerp}}]{westmeier08}
{Westmeier}, T., {Br{\"u}ns}, C., \& {Kerp}, J. 2008, MNRAS, 390, 1691

\end{thebibliography}

\newpage
\onecolumn

\end{document}